\RequirePackage{snapshot}            
\documentclass[letterpaper,twoside]{article}
\usepackage{latexsym,times}
\usepackage[intlimits]{amsmath}
\usepackage{amsfonts,amssymb}
\DeclareSymbolFontAlphabet{\mathbb}{AMSb}
\usepackage[mathscr]{eucal}
\usepackage[notoday]{rcsinfo}
\usepackage{changebar} 
\usepackage[textstyle,cdot,squaren,thickspace,thickqspace]{SIunits}
\usepackage[colon,authoryear]{natbib}
\usepackage{fancyvrb}
\usepackage{url}
\usepackage{xspace}
\usepackage[usenames,dvips]{color}
\usepackage[dvips]{graphicx}
\DeclareGraphicsExtensions{.eps}   
\graphicspath{
  {./}
}
\usepackage[letterpaper,top=1in,bottom=1in,hmargin={1in,1in}]{geometry}
\numberwithin{equation}{section}

\makeatletter
\DeclareRobustCommand\em
{\@nomath\em \ifdim \fontdimen\@ne\font >\z@
  \upshape \else \slshape \fi}

\makeatother

\begin{document}
\title{Fuzzy Cognitive Maps Of Local People Impacted By Dam Construction: Their Demands Regarding Resettlement}
\author{Uygar \"{O}zesmi}
\date{}
\maketitle

\begin{center}Department of Environmental Engineering, Erciyes University, 38039 Kayseri, Turkey, Uygar@ozesmi.org
\end{center}

\rcsInfo $Id: template.tex,v 1.3 2004/06/30 23:26:18 jonnyreb Exp $
\typeout{}
\typeout{<==\rcsInfoFile\ \rcsInfoRevision\ \rcsInfoDate\ \rcsInfoOwner==>}
\typeout{}
\begin{abstract}
Fuzzy cognitive mapping was used to understand the wants and desires of local people before resettlement. Variables that the affected people think will increase their welfare during and after dam construction were determined.  Simulations were done with their cumulative social cognitive map to determine which policy options would most increase their welfare.  The construction of roads, job opportunities, advance payment of condemnation value, and schools are central variables that had the most effect on increasing people's income and welfare. The synergistic effects of variables demonstrated that the implementation of different policies not only add cumulatively to the people's welfare but also have an increased effect.
\end{abstract}


\section*{INTRODUCTION}

Half of the 192 dam projects, which were financed by the World Bank before 1994, had no resettlement plan (WCD, 2000).  In 70 percent of the ones that had a plan there was only financial compensation.  Of these 75 percent did not comply with the criteria set by the World Bank itself.  The loss of productive capacity was compensated in none.  According to an assessment in all of these projects the social and environmental problems could be averted without affecting their financial viability (WCD, 2000).

When resettlement programs for large-scale dam projects do exist, they usually focus on physical settlements rather than direct and indirect effects on social and economic development. However the socio-economic conditions to provide for the welfare of the locals is at least as or may be more important than physical settlement. Resettlement severs people's geographical, social and economic ties and functions.  The land, the house, the livelihood, and the commons of the locals affected by the dam are endangered or lost.  This reduces the socio-cultural resilience of the people.  The endangerment of economic power and complex systems of livelihood prevents people from fulfilling their needs (WCD, 2000).

Usually the amount of people affected by a dam has always been underestimated and the condemnation money payments have been low. In most projects only condemnation values or single-payment monetary compensations were given.  All payments have usually been late and the amount has been insufficient.  For example, in Turkey in a study commissioned by the World Commission on Dams about Aslantas Dam, the payments made for condemnation were not sufficient to buy the same amount of land.  The same situation was valid in Tarbela dam in Pakistan and Kiambere Dam in Kenya, too (WCD, 2000).
Unfortunately people benefiting from dams are not the local people who are negatively impacted by the dam. Especially the most impacted local people are the ones who do not have much capital, are poor, and are marginalized. Local people are the ones who are asked last whether they want the dam or not.  The importance of resettlement policies and their implementation is emphasized, as this is a major factor that determines whether local people will risk resistance or risk resettlement (Dwivedi, 1999).

This study resulted from a more progressive approach that was taken by the Turkish State Hydraulic Works to ask the local people about their demands before dam construction. I hope that this approach will become the norm in such large state projects and will lead to policies that implement these demands at the end. A 540 MW dam is to be built on the Coruh River in northeastern Turkey. The dam is to be 270 m high and the reservoir area will be 33 km2.  The reservoir will completely flood the village of Yusufeli and partially or totally flood 17 surrounding villages, and affect approximately 16,000 people.  The average annual household income of these people is approximately 4500 USD.  Over half of them obtain their income from agriculture, which consists of intensive market-oriented garden agriculture on small plots. All of the agricultural land will be left underwater as it is located in the valleys (Sahara Muhendislik, 2001). This study deals with the actions needed so that the local people are not negatively impacted or driven to poverty after the dam construction. I used fuzzy cognitive mapping to determine the demands and desires of the local people who live in settlements that will be directly affected by the Yusufeli Dam.  

\section*{WHAT IS FUZZY COGNITIVE MAPPING?}

Fuzzy cognitive mapping (FCM) is a process where the participant(s) list the factors that they perceive as important to a particular issue and then draw the causal relationships among these factors.  Instead of a dichotomous yes-no (affects or doesn't affect) the fuzzy aspect allows people to weight the causal relationships, which usually cannot be known precisely, but are described as occurring a little, somewhat, a lot, etc.  Anyone can freely draw causal pictures, or cognitive maps, of problems or systems of any kind.  FCM is applicable for analyzing complex problems that involve the interrelationships between various aspects of society such as economics, environment and politics (Nakamura et al., 1982).  FCM is especially appropriate in soft knowledge domains (e.g. political science, military science, history, international relations, organization theory) where both the system concepts/relationships and the meta-system language are fundamentally fuzzy (Kosko, 1986).  For example, FCMs have been used to model organization behavior and job satisfaction (Craiger et al., 1996) and to suggest ways to improve decision making in sport-fisheries management (Radomski and Goeman, 1996).

\subsection*{A brief history of FCM}

Cognitive mapping takes its roots from graph theory put forward by Euler in 1736.  Harary et al. (1965) put this theory into use in quantitative anthropology to make a structural analysis of observations. The structural analyses were done on maps called "digraphs".  Later the political scientist Robert Axelrod (1976) changed these graphs from the subjective interpretations of anthropologists to the observations of individuals.  These graphs, which Axelrod called 'cognitive maps', were coded from text or were first-hand experiences verbally stated by informants. Kosko (1986) opened the way to a more rigorous analysis of cognitive maps through the use of neural network analysis. Maps can be simulated under different scenarios via signal processing which then becomes an auto-associative memory.  He coined the term fuzzy cognitive maps (FCM) for these maps with weighted connections. Carley (1988) put forward some of the statistical techniques so that stakeholders and different groups of people could be compared and contrasted through their cognitive maps.  Cognitive mapping has been used to examine decision making processes and complex social systems (Axelrod, 1976; Bauer, 1975; Bougon et al., 1977; Brown, 1992; Carley and Palmquist, 1992; Cossette and Audet, 1992; Hart 1977; Malone, 1975; Montazemi and Conrath, 1986; Nakamura et al., 1982). These studies opened the way for using cognitive maps to represent local knowledge systems as told by informants (\"{O}zesmi, 1999).

\subsection*{Comparison with other methods}

In this study I use fuzzy cognitive mapping to approach the question of what people impacted from the dam want and need during and after dam construction. Fuzzy cognitive mapping can offer advantages over other social science research methods such as questionnaire surveys, unstructured interviews, or rapid assessment processes (RAP). A disadvantage of questionnaire surveys is that the researcher asks questions that they consider important and also typically give a choice of answers.  In the FCM methodology the interviewee determines what is important and what should be included.  Questionnaire surveys have the advantage that large numbers of people can be surveyed.  In fact, questionnaire surveys ($n = 2522$) were employed as one research method in the Yusufeli Dam resettlement plan (Sahara Muhendislik, 2001). However because of an anticipated low participation in written surveys, the surveys had to be completed in person, and thus took a considerable amount of time on the part of researchers. Although the FCM technique was done with less people than the questionnaire surveys in this study, it provided more insight into the desires and needs of the local people (Sahara Muhendislik, 2001).  However, I am not advocating that only a FCM method should be used in dam resettlement projects.  Detailed information needs to be obtained from the individuals that will be affected about their losses so that they can be adequately compensated, and a FCM method would not be used to obtain this information from all affected individuals.  Rather the FCM becomes a way to obtain people's opinion about the socio-economic conditions that would be most helpful for increasing their welfare during and after resettlement.

Open-ended interviews could be used to obtain this information, but the researchers must extract the important concepts and relationships from the interview transcripts, which is time-consuming and may lead to biases and misjudgments.  In the FCM technique the important concepts and relationships are drawn on the map by the interviewees, thus removing potential researcher bias as well as reducing the amount of time spent on analysis after the interviews.  FCMs can be quickly coded from the drawn maps.  In addition results from open-ended interviews are highly qualitative and do not permit statistical analysis.  With FCM, maps of different stakeholder groups can be compared and the results, while qualitative, do yield relative information as to what is most important or what things have the most effect on a system.
FCM is similar to RAP in that it is a systems approach.  In the case of FCM, the interviewees themselves determine the things affecting the system and how they relate to each other.  It can also be an iterative process, as maps can be obtained from any number of interviewees and these maps can be superimposed to form a social cognitive map or the maps can be superimposed according to stakeholder group to compare differences and similarities among groups of people.  Thus the FCM method also has advantages for enabling stakeholder participation and analyzing stakeholder group maps for participatory methods (\"{O} and \"{O}zesmi, 2001) and also in cases of conflict among stakeholder groups (\"{O}zesmi, 1999).  Certainly FCM could be incorporated as one of the suitable techniques for use in RAP because FCM is especially appropriate for information gathering where the researcher wants to obtain from the local people what they think is important. In FCM, the question can be and usually is very general, such as, "what are the important factors/variables affecting the system and how do these factors affect each other?"

\section*{METHODS}

A standardized cognitive mapping methodology was used in this research to bring to light the wants and desires of the population impacted from the dam. The analysis of data has been done through graph theory and fuzzy cognitive mapping, a form of signal processing analysis using neural network simulations (\"{O}zesmi, 1999).  

\subsection*{Field Research}

Interviews were made in settlements within the four geographic areas that will be flooded by the dam lake.  The elected village heads and their spouses were called by telephone one or two days ahead of the interview.  An appointment was made to meet with the village head and their spouses for 1 to 1.5 hours to make a graphical exercise about the dam to be built (permission for an appointment was not refused by informants except for an NGO at the Yusufeli township).  A total of 13 cognitive maps were drawn with village headmen and their spouses, and an NGO head and his spouse.  The total number of people who participated in this research was 26 people.  The exercise was done with married couples, so that participation of both genders could be secured.  Women in general made different contributions to maps at about 30 per cent, and have assured us that they agree with the contributions of their husbands.  Men have likewise agreed on the contributions of women.  Thus I believe I have represented the concerns of the women sufficiently in these maps.
 
Before starting the cognitive mapping exercise I explained the state of the building of the dam and why this research was being made.  Then I explained what kind of graph we were going to draw using a completely unrelated map that summarized the occurrence of traffic accidents. On the corner of a large piece of paper (50X30 cm) the location, date, time and duration, and the names of the participants, their gender, age and occupation were noted (Figure 1).  Then the two questions relating to the listing of variables and the drawing of the graph were asked and written on top of the paper. These questions were: "What are the variables, factors, causes and things that are going to increase or decrease the welfare of the people after the dam construction during resettlement?" and "How do these variables, factors, causes and things affect each other?"  As the interviewees stated these variables, they were listed on the left of the paper.  We drew the variable 'peoples' welfare" in the middle with the rest of the variables drawn around it.  Then we drew the causal relationships among the variables.  The causal relationships were given by the participants as either positive (+) or negative (-) and were weighted by them as strong (1), medium (0.5) or weak (0.25) (Figure 1).

\subsection*{Data Analysis}

Variables were listed and pooled cumulative variable richness plots were drawn to find out whether variable (sampling) saturation was reached or not.  For this purpose Monte-Carlo simulations were done using random numbers from Minitab 10.5 statistical package to determine the confidence interval of observed saturation.

The cognitive maps drawn by informants were coded as neighborhood matrices. Indices characterizing the structure and type of variables were calculated by cell programming. A macro program written in the spreadsheet was used to make neural network simulations. Neural network simulation is done using matrix algebra. The state vector of variables is multiplied with the neighborhood matrix and then the results are squashed into the interval of [1,0] by a logistic function.  The result is the new vector-state. The multiplication is continued until the system stabilizes into one variable, falls into an oscillation or goes into chaos.  All the simulations ended in a stable state.  Policies can be simulated by clamping the vector-state of a variable to a desired value while doing the multiplication process.  The social cognitive map is obtained by matrix augmentation in which each participant map is given an equal weight. Details with regards to the above mentioned methodology can be found in \"{O}zesmi (1999) or at this website address: http://env.erciyes.edu.tr/Kizilirmak/UODissertation.html.

\section*{RESULTS}

\subsection*{Saturation of Collected Data}

Cognitive mapping exercises in the area to be affected by the dam took an average of $56 \pm 15$ SD minutes (min 35, max 95 min.).  The increase in new variables considerably declined at the end of 13 cognitive maps and additional new variables decreased to 3 per new map (Figure 2).  Since these new additions were very much personal and special issues to the people interviewed I believe the results indicate the wants and desires of the majority of the people in the region affected by the dam.

\subsection*{The Structure and Characteristic of Cognitive Maps Obtained}

\subsubsection*{Individual cognitive maps}

The results obtained with the maps from the settlements affected by the dam are comparable to previous studies in Turkey (Table 1) and other places.  In a study of a small business environment, 57 concepts and 87 links were mentioned by the owner , which resulted in a connection to variable ratio of 1.53 and a density of 0.027.  A study with a modal average of 75 minutes of interviewing of 116 informants elicited 32 concepts in the range of 14-59 .  Eden et al.  found typical 1.15-1.20 ratios of links to nodes. I found $1.64 \pm 0.95$ SD ratio of connections to variables in the Kizilirmak Delta, which is almost the same as Yusufeli with $1.76 \pm 0.51$ SD (Table 1).

The many transmitter variables mentioned by Yusufeli people show that they see many factors beyond their control affecting them.  The very few receiver variables other than 'the welfare of people' means that they do not see much utility in the existing systems and that the prospects are limited. Their maps are not hierarchical.  Variables are affecting each other sometimes in loops. The non-hierarchical maps are also a result of the high connectivity of the maps.  The high connectivity is an indication of the interdependent character of the mentioned variables (Table 2). 

\subsubsection*{Social cognitive maps}

In this study the number of variables and connections were again comparable to previous studies similar to the individual cognitive maps.  Carley and Palmquist  reported that 29 undergraduates mentioned up to 244 concepts on research writing, and 45 students produced 217 concepts on tutor selection. One individual could only consider between 30 and 40 concepts in a session lasting between 20-40 minutes because of combinatorial explosion.  Nakamura et al.  obtained 152 concepts and 265 connections from 5 documents on traffic problems in Japan.  \"{O}zesmi (1999) obtained 136 variables and 616 connections from 31 cognitive maps that included 3 distinctly different stakeholder groups in perceptions of a wetland ecosystem in Turkey. In this study I found 97 variables and 360 connections in 13 maps of 26 people (Table 1).  

\subsection*{Variables Mentioned and Their Importance}

There are three different ways of at looking at different variables mentioned in a cognitive map.  Variables can be ordered first by how many times they were mentioned in each map, second by their centrality (including the components of how much input they receive, \underline{indergree}, and how much output they give, \underline{outdegree}), and third by the type of these central variables (\underline{transmitter}, \underline{ordinary}, \underline{receiver}).  The variable type reveals how people think about the variables.

The importance of variables to people who are going to be affected by the dam can be represented by how many times they have been mentioned in cognitive maps (Table 3).  School was mentioned in all 13 cognitive maps.  Construction of roads, job opportunities, and advance payment of condemnation value were all mentioned in 12 out of 13 maps.  Ordering variables by the number of times they have been mentioned gives an order of importance in the eyes of the people who are going to be affected by the dam.  They indicate what they would like to see happen after or while the dam is constructed.

More important than the repetition of variables is to find out how these variables are in relation to others. Centrality and other indices can be used to evaluate the structure of the cognitive map.  The centrality of a variable shows how connected it is to other variables and the cumulative strength of these variables.  The centrality is composed of the indegree and of the outdegree.  If the indegree is zero and there is a non-zero outdegree the variable is called a transmitter, if in the opposite it has a non-zero indegree and a zero outdegree then it is a receiver variable.  If the variable has both a non-zero indegree and outdegree than it is called an ordinary variable.  Of course ordinary variables based on the ratio of their indegrees and outdegrees can be more or less receiver or transmitter variable.

The centrality of the variable is therefore not only a frequency of expression but also how important that given variable is given the whole structure of the cognitive map.  The type of variables, their in and outdegrees and their centrality for the social cognitive map of the people affected by the dam, are given in Table 4.  Variables that have a centrality of 0.5 and lower are not included in Table 4 for brevity.  After people's welfare, the most central variables were job opportunities, construction of roads, schools, government support, and industrial facilities, all with a centrality > 2.0. 

\subsection*{Cognitive Map Simulations}

The real advantage of cognitive mapping is to be able to evaluate different policies and judge their outcome to the people's welfare based entirely on the people's cognitive map without any interference and judgement imposed by authorities or scientists. The policies, the outcome variables, and the causal relationship between variables are entirely that of the informants.  What remains afterwards is the implementation of the wants and desires of the people after dam construction.  The most important variable, which needs to be increased after and during dam construction, is the welfare of the people.  The effect of the most mentioned or central single policies and the synergetic effects of these different policies were simulated to maximize the welfare of people.

A relative increase over the steady state of variables when no policies are implemented is used to compare different policies mentioned by informants (Figure 3 and 4).  These graphs are the result of simulations as described in the data analysis section. The variables (policies implemented) in the graphs are ordered from the most effective to the least.  The synergistic effects of variables that are most central are simulated in pairs and in triples to demonstrate that the implementation of different policies do not only add cumulatively to the welfare of the people but have an increased effect (Figure 5 and 6). 

As all policies are implemented a saturation of effect will be observed (Figure 7).  The most central variables will add a considerable amount to the welfare. As other policies are added the total effect of policies to welfare will slow down, but is nevertheless important.  It is not a surprise given Figure 3, 5 and 6 that the most central variables will add more to the welfare of people and are important to address, since these are variables that people most concur on.  

These simulations help to pinpoint the most important issues to be addressed by government agencies after and during dam construction. 

\section*{DISCUSSION AND CONCLUSION}

Looking at the mentioned variables, education is the most mentioned variable that provides for economic independence and welfare (Table 3).  All the interviewees had education in their maps.  The reconstruction of schools, which are left underwater, will help local economy and the construction of new schools is going to increase the wealth of the area.  Informants think that students coming from other places in the country, such as the ones in state boarding schools, liven up local economies and create an education economy.  Schools are also the third most central variable (Table 4, excluding people's welfare, which we are trying to maximize) and one of the most important contributors to welfare (Figure 3 and 7).

	The second most mentioned variable, construction of roads (Table 3), was a very prominent concern that came up in every discussion.  Construction of roads was also the second most central variable (Table 4) and had a stronger outdegree making it closer to a transmitter variable, hence a forcing function. It was also the second most important variable contributing to the people's welfare (Figure 3 and 7).  It is one of the variables with highest synergistic effect on welfare (Figure 5 and 6). The village headmen demand that the roads be constructed in parallel to the dam construction and be finished contemporaneously with the dam.

The people of Yusufeli are very afraid that they are going to loose their jobs and want to secure a future for their children, therefore job opportunities is also a frequently mentioned variable (Table 3) and it is the most central variable.  According to informants a lot of other variables depend on having a job.  In addition, if other variables are addressed then more jobs will be created, because the indegree of job opportunities is higher than its outdegree (Table 4).  Job opportunities will not only have a significant impact on welfare but it is also one of the few variables that do influence the income of the people directly (Figure 3 and 4).  Other variables that impact the income of the people in order of contribution are advance payment of condemnation values, revenues from the dam, a lively city center and government support (Figure 4).  The advance payment of condemnation values was another issue that everybody raised during the investigator's visit to the area.  It is not only considered to have the largest contribution to income but also to the welfare of the people.  The payment for loss due to condemnation must be just and timely, preferably more than the actual worth of the land.  Implementing such a policy will help people impacted by the dam to generate more income and increase their welfare.  

The construction of the roads, job opportunities and the advance payment of the condemnation value are issues that every dam project has to take seriously into account.  These three issues have also come up in almost all the cognitive maps as some of the most important variables to affect the welfare of the people.  The inhabitants do not think that these issues are going to be resolved easily. Infrastructure and government support to recreate the previously existing facilities is again among the most mentioned and most central values (Table 3 and 4), they also contribute directly (Table 5 and 9) and have synergistic effects in increasing the welfare of people with other variables (Table 7 and 8).  In countries such as China, Thailand, and Malaysia the roads, electricity, water and school facilities lost to the dam was never replaced, or it was constructed so late that people dispersed slowly leaving no trace behind (WCD, 2000). 

In China the government reports on dams mention "the seven difficulties" of resettlement projects.  These are electricity, drinking water, schools, food security, health services, telecommunication and transportation (WCD, 2000).  The people of Yusufeli have seen all these difficulties without any knowledge of previous dams and have put them into their cognitive maps frequently and with high centralities (Tables 3 and 4).  These variables have important consequences for their welfare (Figures 3-7).

Again according to the cognitive maps, informants state that the people of Yusufeli want to be able to work in the dam during construction but also afterwards as maintenance and running staff.  

Even if all these policies are implemented the misfortune of the people who lost their livelihood, their property and memories under water cannot prevented. There are two important variables mentioned in the maps that have been addressed in the world and can be addressed in the case of Yusufeli. The first is revenues from the dam, which is also the second most important variable to generate income according to simulations (Figure 4).  The second is inexpensive or free electricity.  There are many existing examples from the world and there is no reason why these could not be implemented in the Yusufeli dam resettlement project.

In Norway there is an example where either a special price for electricity or free electricity is given to displaced people.  There are revenue sharing programs in Urra 1 Dam in Colombia and in the cross-boundary Itaipu project in both Brazil and Paraguay.  There is also revenue sharing in Minastuck project as well as Hydro-Quebec projects in Canada (WCD, 2000).  In the world energy sector it is common practice that a certain percentage of proceeds is given back to the local communities or municipalities.  

The main goal of any resettlement project must be to increase of the welfare of the people who are going to be undoubtedly negatively impacted by the dam after and during its construction.  A dam, being decided somewhere in the state apparatus, is an imposition on the local people.  Therefore, at least to decide how to help the same people, it is wiser to ask them first, rather than deciding top down on what the best way to increase their welfare is.  Through cognitive mapping one can best approach the wants and desires of the people through their own way of thinking in a bottom-up way.

With fuzzy cognitive mapping the people themselves decide what variables are important and how much they would affect their welfare.  The maps are easy for people to draw.  It is simple to code the maps and to augment them into one social cognitive map.  Any number of maps can be combined; the maps can be of different sizes and with different variables and causal connections.  By looking at the most mentioned variables and most central variables it is clear which variables are most important for the people as a group.  Finally simulations can be done quickly and easily to see the effect of different policy options.


\section*{REFERENCES}

\noindent
Axelrod, R. (1976) \textit{Structure of Decision, The Cognitive Maps of Political Elites.} Princeton, NJ: Princeton University Press.

\noindent
Bauer, V. (1975) Simulation, Evaluation and Conflict Analysis in Urban Planning, in: M. M. Baldwin (ed.) \textit{Portraits of Complexity: Applications of Systems Methodologies to Societal Problems}, p. 119-126. Columbus, OH: Batelle Institute.

\noindent
Bougon, M., Weick, K., and D. Binkhorst (1977) Cognition in Organizations: An Analysis of the Utrecht Jazz Orchestra, \textit{Administrative Science Quarterly} 22:606-639.

\noindent
Brown, S. M. (1992) Cognitive Mapping and Repertory Grids for Qualitative Survey Research: Some Comparative Observations, \textit{Journal of Management Studies} 29:287-307.

\noindent
Carley, K. (1988) Formalizing Social Expert's Knowledge, \textit{Sociological Methods and Research} 17:165-232.

\noindent
Carley, K., and M. Palmquist (1992) Extracting, Representing, and Analyzing Mental Models, \textit{Social} Forces 70:601-636.

\noindent
Cossette, P., and M. Audet (1992) Mapping of an Idiosyncratic Schema, \textit{Journal of Management Studies} 29:325-347.

\noindent
Craiger, J.P., Weiss, R.J., Goodman, D.F., and A.A. Butler (1996) Simulating Organizational Behavior with Fuzzy Cognitive Maps, \textit{International Journal of Computational Intelligence and Organizations} 1:120-133.

\noindent
Dwivedi, R. (1999) Displacement, Risks and Resistance: Local Perceptions and Actions in the Sardar Sarovar, \textit{Development and Change} 30:43-78.

\noindent
Eden, C., F. Ackerman, and S. Cropper (1992) The Analysis of Cause Maps, \textit{Journal of Management Studies} 29:309-323.

\noindent
Harary, F., R. Z. Norman, and D. Cartwright (1965) \textit{Structural Models: An Introduction to the Theory of Directed Graphs}. New York, NY: John Wiley \& Sons.

\noindent
Hart, J.A. (1977) Cognitive Maps of Three Latin American Policy Makers, \textit{World Politics} 30:115-140.

\noindent
Kosko, B. (1986) Fuzzy cognitive maps, \textit{International Journal of Man-Machine Studies} 1:65-75.

\noindent
Malone, D.W. (1975) An Introduction to the Application of Interpretive Structural Modeling, in: M. M. Baldwin (ed.) \textit{Portraits of Complexity: Applications of Systems Methodologies to Societal Problems}, p. 119-126. Columbus, OH: Batelle Institute.

\noindent
Montazemi, A.R., and D.W. Conrath (1986) The Use of Cognitive Mapping for Information Requirements Analysis, \textit{MIS Quarterly} 10:45-55.

\noindent
Nakamura, K., S. Iwai, and T. Sawaragi (1982) Decision Support Using Causation Knowledge Base, \textit{IEEE Transactions on Systems, Man, and Cybernetics} SMC-12:765-777.

\noindent
\"{O}zesmi, U. (1999) \textit{Conservation Strategies for Sustainable Resource Use in the Kizilirmak Delta in Turkey}, Ph.D. Dissertation, University of Minnesota, St. Paul, 230 pp.

\noindent
\"{O}zesmi, U. and S. \"{O}zesmi (2001)  A Participatory Approach to Ecosystem Conservation: Uluabat Lake Environmental Management Plan Using Fuzzy Cognitive Maps and Stakeholder Analysis, \textit{Proceedings of the IV.National Environmental Engineering Congress}, Mersin, Turkey, (7-10 November), pp. 16-24.

\noindent
Radomski, P.J., and T.J. Goeman (1996) Decision Making and Modeling in Freshwater Sport-fisheries Management, \textit{Fisheries} 21:14-21.

\noindent
Sahara Muhendislik Ltd. STI.  (2001) \textit{Yusufeli Baraji Yeniden Yerlisim Plani Sonuc Raporu} (in Turkish, Final Report on the Resettlement Plan for Yusufeli Dam), 287 pp., Ankara: Sahara Muhendislik Ltd. STI. 

\noindent
World Commission on Dams (2000) \textit{Dams and Development: A New framework For Decision-making}. London: Earthscan Publications.

\clearpage

Uygar \"{O}zesmi was a Fulbright Scholar at The Ohio State University, where he completed a M.S. in Environmental Science, and MacArthur Scholar at the University of Minnesota, where he completed a Ph.D. in Conservation Biology with a minor in Development and Social Change.  He is currently assistant professor and Chair of Environmental Science, Department of Environmental Engineering, Erciyes University, 38039 Kayseri, Turkey (email uozesmi@erciyes.edu.tr).  His research interests include sustainable development and social change, participatory processes and ecosystem conservation, ecosystem modelling, sustainable livelihoods, environmental NGO capacity building, artificial neural networks, and fuzzy cognitive mapping.  

\clearpage

\linespread{1}

\begin{table}[ht]
\begin{center}
\caption{The structural indices of individual cognitive maps of Yusufeli as compared to the Kizilirmak Delta in Turkey.}
\begin{tabular}{l|p{0.8in}|p{0.3in}|p{0.65in}|p{0.8in}|p{0.3in}|p{0.65in}}
\hline
&Individual Cognitive Map Mean ($n=31$)&SD&Social Cognitive Map ($n=31$)&Individual Cognitive Map Mean ($n=13$)&SD&Social Cognitive Map ($n=13$)\\
\hline
Number of Variables&19&7&136&24.2&5.7&97\\
Number of Transmitter Variables&7.0&4.3&27&9.1&3.2&24\\
Number of Receiver Variables&3.0&2.4&9&1.6&1.1&2\\
Number of Ordinary Variables&8.9&3.3&100&13.5&6.5&71\\
Number of Connections&28.3&10.6&616&43.6&17.6&360\\
Connection/Variable&1.64&0.95&4.616&1.76&0.51&3.711\\
Density&0.112&0.109&0.033&0.078&0.022&0.038\\
Hierarchy&0.082&0.135&0.026&0.025&0.014&0.000\\
\hline
\end{tabular}
\end{center}
\end{table}

\clearpage

\begin{table}[ht]
\begin{center}
\caption{The structural indices of individual cognitive maps drawn by husband and wife teams in settlements, which are going to be impacted by the Yusufeli Dam.}
\begin{tabular}{p{1.4in}|p{0.27in}|p{0.25in}|p{0.27in}|p{0.27in}|p{0.25in}|p{0.25in}|p{0.27in}|p{0.25in}|p{0.25in}|p{0.25in}|p{0.25in}|p{0.38in}|p{0.33in}}
&Irmak- yani&Yeni- koy&Kinali- cam&Mor- kaya&Tek- kale&Cev- reli&Celtik duzu&Kilic kaya&Alan- basi&Bah- celi&Dere- ici&Yusufeli Merkez&Yusufeli Dernegi\\
\hline
\textbf{\# of Variables}&21&29&21&29&15&22&28&24&23&19&18&30&35\\
\textbf{\# of Transmitter Variables}&10&12&9&4&9&3&11&10&4&14&7&6&9\\
\textbf{\# of Receiver Variables}&2&2&3&4&1&0&1&1&1&1&1&1&3\\
\textbf{\# of Ordinary Variables}&9&15&9&21&5&9&16&13&18&4&10&23&23\\
\textbf{\# of Connections}&30&50&29&50&16&51&45&45&67&24&25&68&67\\
\textbf{Connection/Variable}&1.429&1.724&1.381&1.724&1.067&2.318&1.607&1.875&2.913&1.263&1.389&2.267&1.914\\
\textbf{Density}&0.071&0.062&0.069&0.062&0.076&0.110&0.060&0.082&0.132&0.070&0.082&0.078&0.056\\
\textbf{Hierarchy}&0.020&0.027&0.027&0.018&0.012&0.030&0.016&0.016&0.057&0.011&0.040&0.044&0.011\\
\hline
\end{tabular}
\end{center}
\end{table}

\clearpage

\begin{table}[ht]
\begin{center}
\caption{Variables mentioned in 3 or more out of 13 total cognitive maps and the number of maps where they were mentioned.}
\begin{tabular}{l|l}
\textbf{Variables}&\textbf{Repeat}\\
\hline
Peoples Welfare&13\\
School&13\\
Construction of Roads&12\\
Job Opportunities&12\\
Advance payment of condemnation value&12\\
Infrastructure&10\\
Tailoring&10\\
Health Services&10\\
Industrial facilities&9\\
Vocational Education&8\\
Carpet Weaving&7\\
Animal Husbandry&7\\
Agriculture&7\\
Revenue from the dam&6\\
Government support&6\\
Culture and knowledge&6\\
Irrigation&6\\
Working in the dam&5\\
Constructing the dam shortly&5\\
Mosque&5\\
Drinking water&5\\
Closeness of the city&5\\
Resettlement somewhere in the West&4\\
Computer use&4\\
Giving electricity inexpensive&4\\
Having the village in one location&4\\
Telephone&4\\
Development of tourism&4\\
Being able to stay where they are now&4\\
Construction of the resettlement buildings&4\\
Yusufeli identity&4\\
Homes with gardens that address needs&4\\
A lively town center&4\\
Land given out by the treasury for resettlement&3\\
Resettlement in the mountain commons&3\\
Population&3\\
Marketing&3\\
Resettlement in one single county&3\\
The suitability of the town for resettlement&3\\
\hline
\end{tabular}
\end{center}
\end{table}

\clearpage

\begin{table}[ht]
\begin{center}
\caption{The variables ordered by their centrality, their indegree and outdegree and what type of variable they are based on their indegree and outdegree.}
\begin{tabular}{l|llll}
&Variable Type&Indegree&Outdegree&Centrality\\
\hline
Peoples welfare&Ordinary&12.44&0.12&\textbf{12.56}\\
Job opportunities&Ordinary&2.92&1.23&\textbf{4.15}\\
Construction of the roads&Ordinary&0.65&2.85&\textbf{3.50}\\
Schools&Ordinary&0.98&1.87&\textbf{2.85}\\
Government support&Ordinary&0.37&1.94&\textbf{2.31}\\
Industrial facilities&Ordinary&0.58&1.60&\textbf{2.17}\\
Infrastructure&Ordinary&0.56&1.38&\textbf{1.94}\\
Advance payment of condemnation value&Ordinary&0.19&1.65&\textbf{1.85}\\
Agriculture&Ordinary&1.19&0.46&\textbf{1.65}\\
Vocational education&Ordinary&0.31&1.21&\textbf{1.52}\\
Being able to stay where they are now&Ordinary&1.23&0.27&\textbf{1.50}\\
Animal husbandry&Ordinary&0.85&0.58&\textbf{1.42}\\
Health services&Ordinary&0.79&0.62&\textbf{1.40}\\
Culture and knowledge&Ordinary&1.00&0.33&\textbf{1.33}\\
A lively town center&Ordinary&0.52&0.73&\textbf{1.25}\\
Population&Ordinary&1.00&0.19&\textbf{1.19}\\
Tailoring&Ordinary&0.60&0.54&\textbf{1.13}\\
Resettlement somewhere in the West&Ordinary&0.38&0.69&\textbf{1.08}\\
Drinking water&Ordinary&0.46&0.52&\textbf{0.98}\\
Constructing the dam shortly&Transmitter&0.00&0.92&\textbf{0.92}\\
Resettlement within the boundaries of Yusufeli&Ordinary&0.31&0.58&\textbf{0.88}\\
Construction of the resettlement buildings&Ordinary&0.27&0.58&\textbf{0.85}\\
Revenue from the dam&Ordinary&0.12&0.69&\textbf{0.81}\\
Carpet weaving&Ordinary&0.37&0.42&\textbf{0.79}\\
Working in the dam&Ordinary&0.23&0.50&\textbf{0.73}\\
Closeness of the city&Ordinary&0.15&0.58&\textbf{0.73}\\
Computer use&Ordinary&0.19&0.50&\textbf{0.69}\\
Togetherness and unity&Ordinary&0.33&0.33&\textbf{0.65}\\
Having the village in one location&Ordinary&0.23&0.40&\textbf{0.63}\\
Development of tourism&Ordinary&0.31&0.31&\textbf{0.62}\\
Resettlement in the mountain commons&Ordinary&0.31&0.27&\textbf{0.58}\\
Up-to-date technology&Ordinary&0.23&0.35&\textbf{0.58}\\
Income&Ordinary&0.50&0.08&\textbf{0.58}\\
Development priority area designation&Ordinary&0.08&0.50&\textbf{0.58}\\
Irrigation&Ordinary&0.15&0.38&\textbf{0.54}\\
The suitability of the town for resettlement&Ordinary&0.15&0.38&\textbf{0.54}\\
\hline
\end{tabular}
\end{center}
\end{table}

\clearpage

\begin{figure}[ht]
\begin{center}
\includegraphics[width=\textwidth]{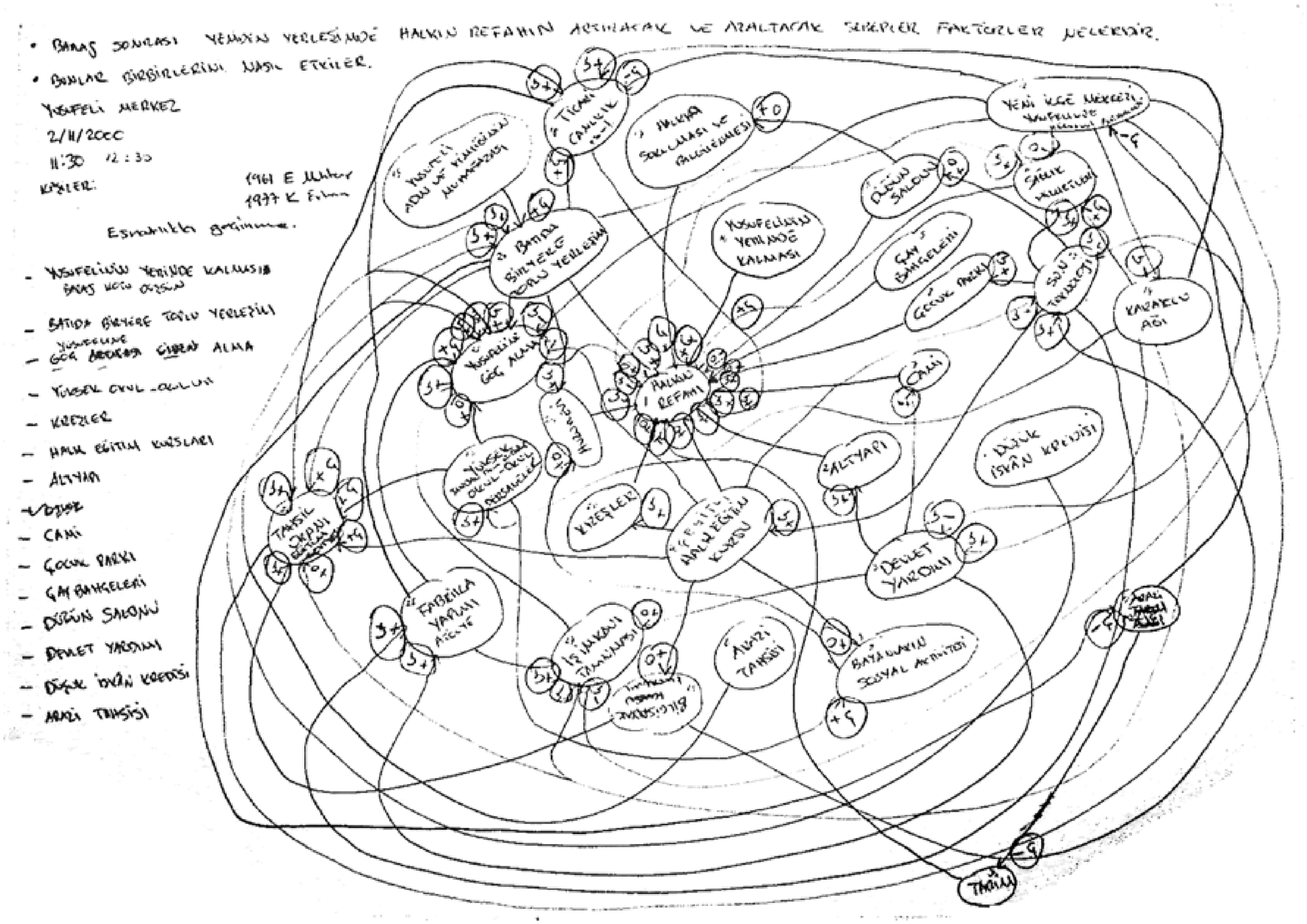}
\caption{A cognitive map drawn by a husband and wife team.}
\end{center}
\end{figure}

\clearpage

\begin{figure}[ht]
\begin{center}
\includegraphics[width=\textwidth]{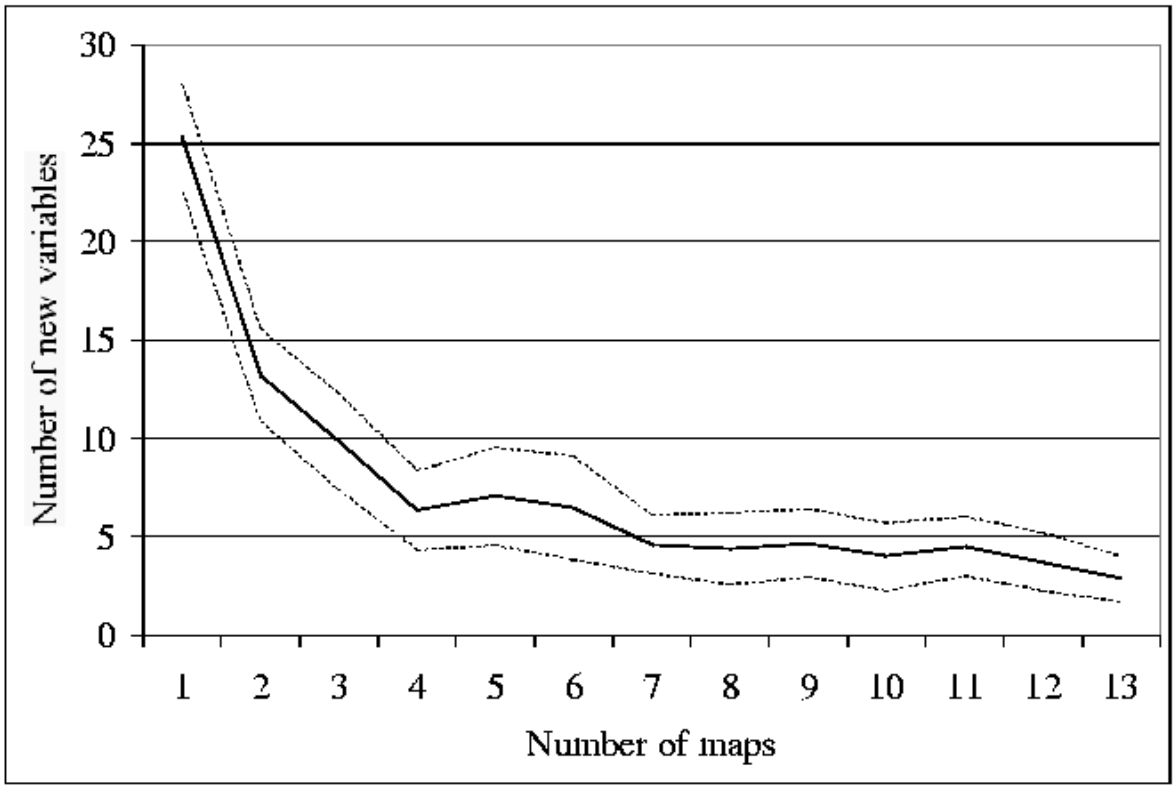}
\caption{The continuously decreasing number of new variables by increasing sample size (center line is the average calculated by 15 Monte-Carlo simulations, and the dashed lines show the 95per cent confidence interval).}
\end{center}
\end{figure}

\clearpage

\begin{figure}[ht]
\begin{center}
\includegraphics[width=\textwidth]{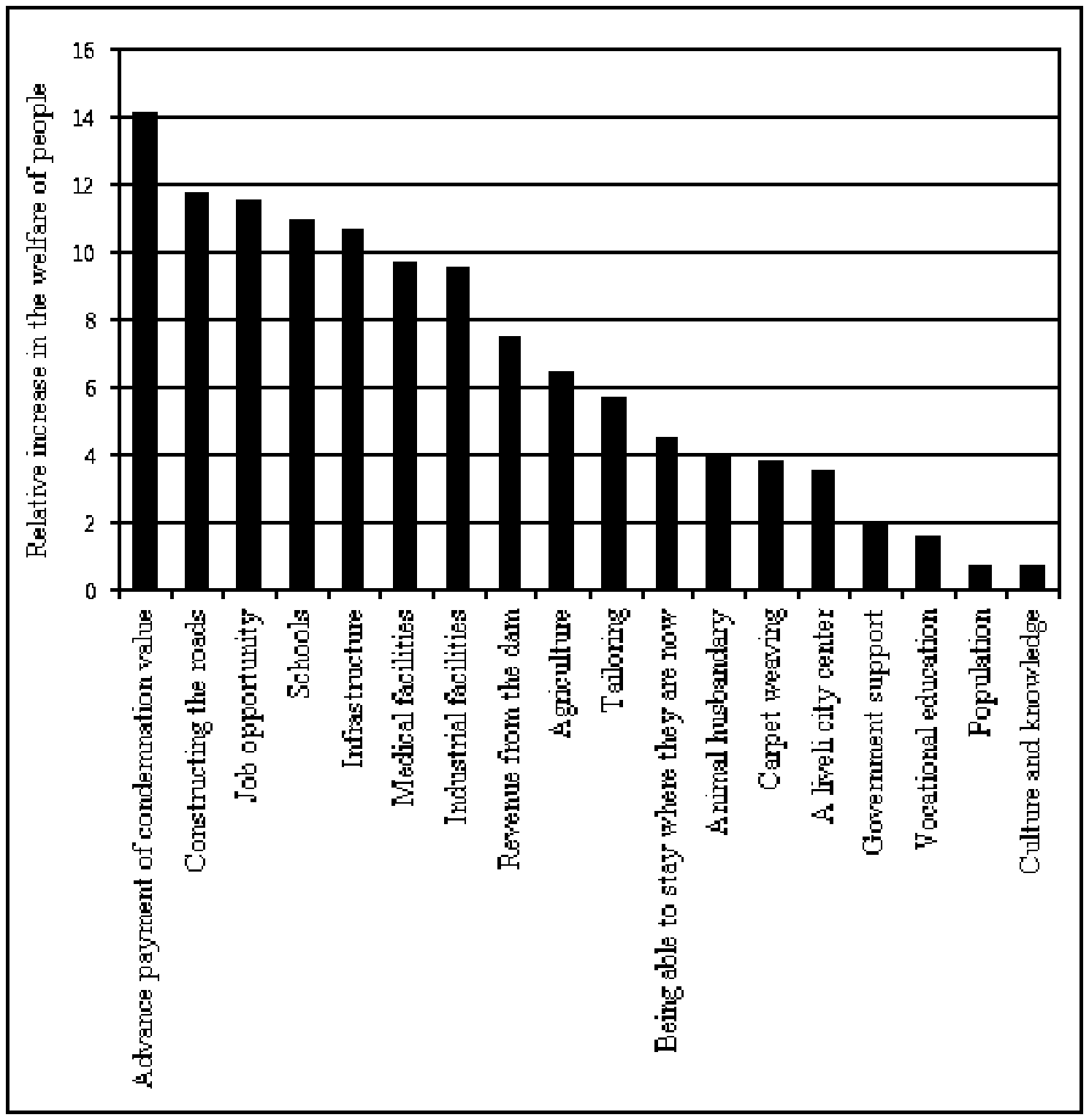}
\caption{The effect of the most central variables causing a relative increase in the welfare of the people impacted by the dam construction.}
\end{center}
\end{figure}

\clearpage

\begin{figure}[ht]
\begin{center}
\includegraphics[width=\textwidth]{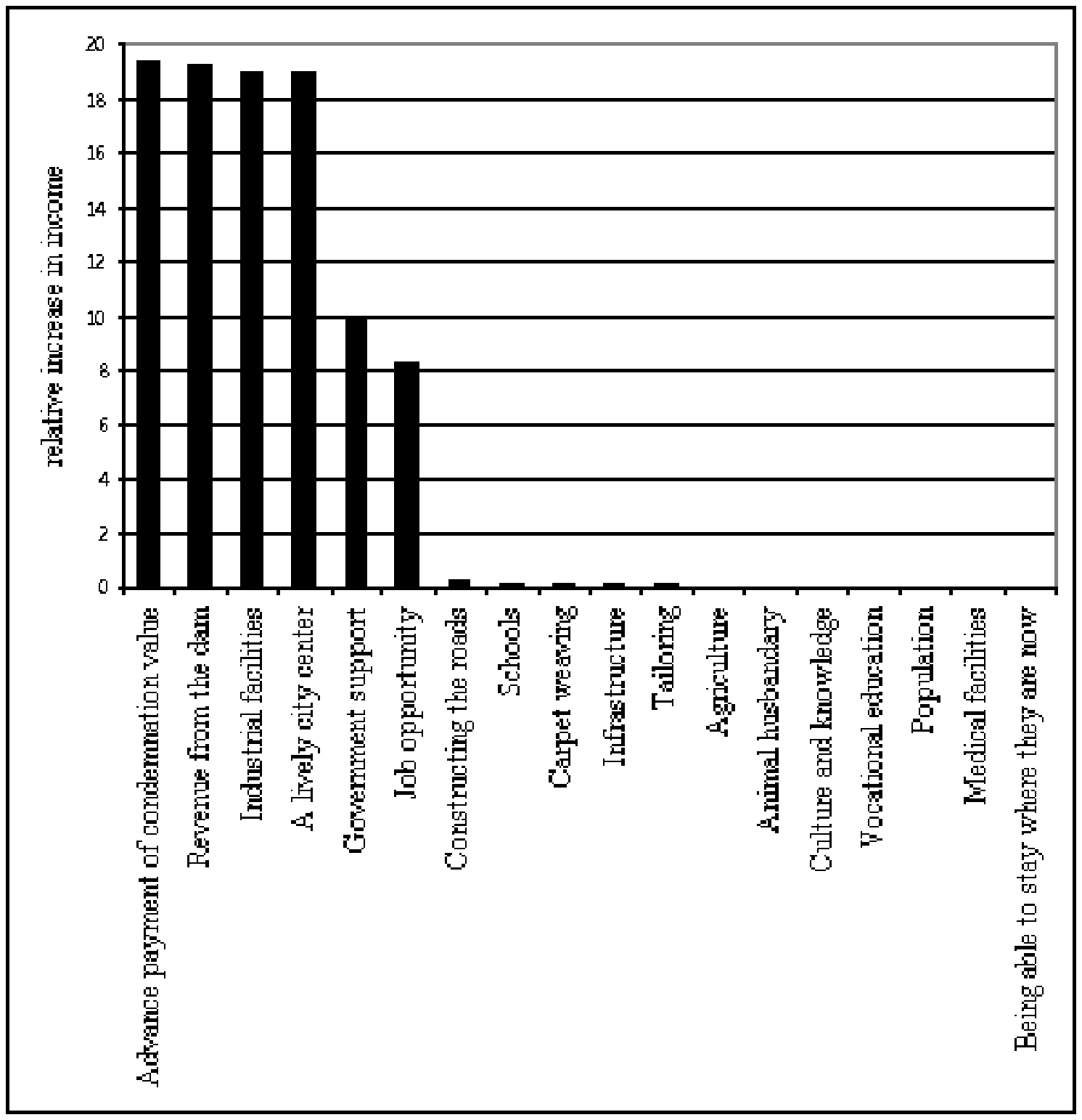}
\caption{The effect of the most central variables causing a relative increase in the income of the people impacted by the dam construction.}
\end{center}
\end{figure}

\clearpage

\begin{figure}[ht]
\begin{center}
\includegraphics[height=7.5in]{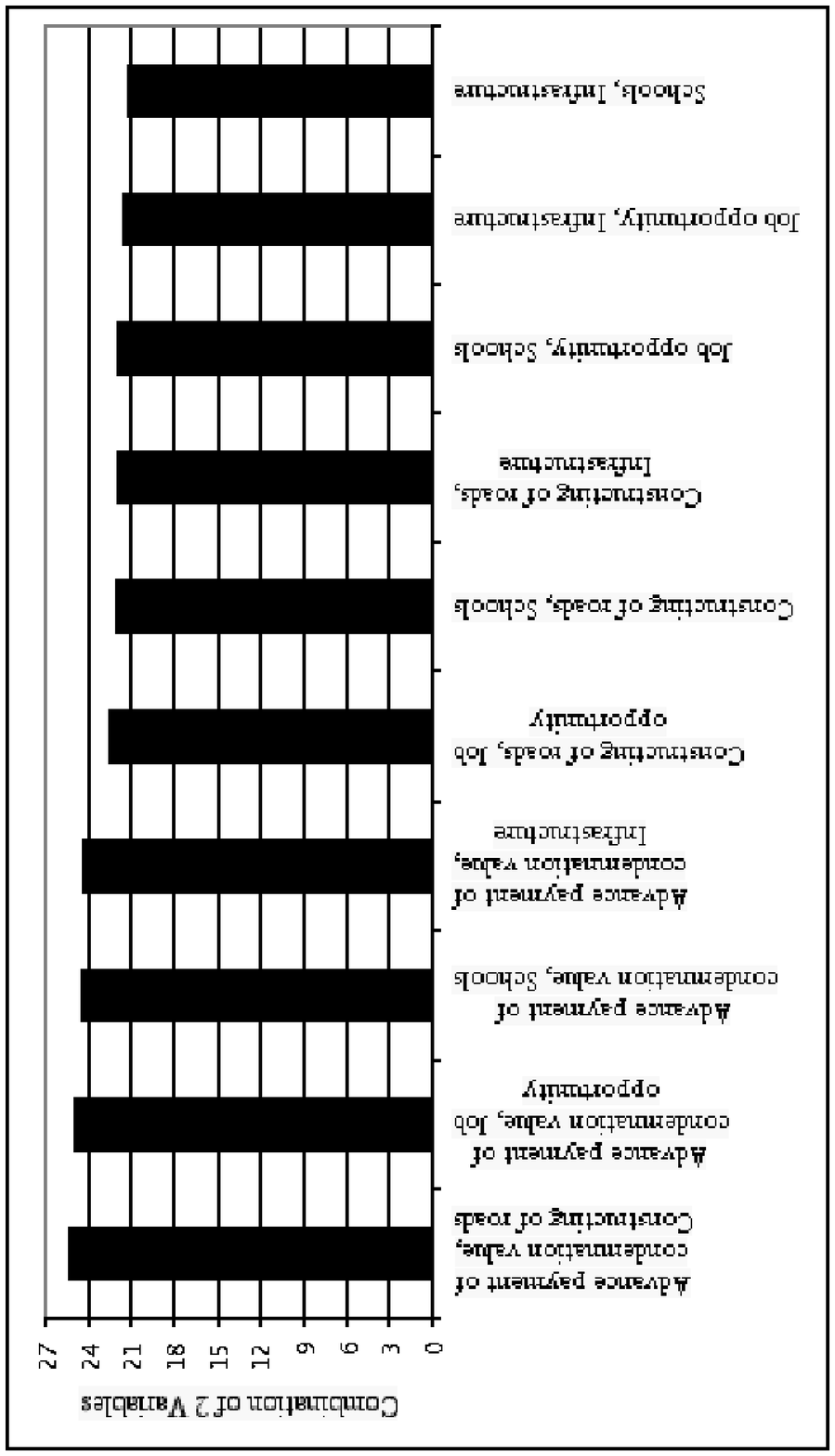}
\caption{The synergetic effect of the most central variables in pairs causing a relative increase in the welfare of the people impacted by the dam construction.}
\end{center}
\end{figure}

\clearpage

\begin{figure}[ht]
\begin{center}
\includegraphics[width=\textwidth]{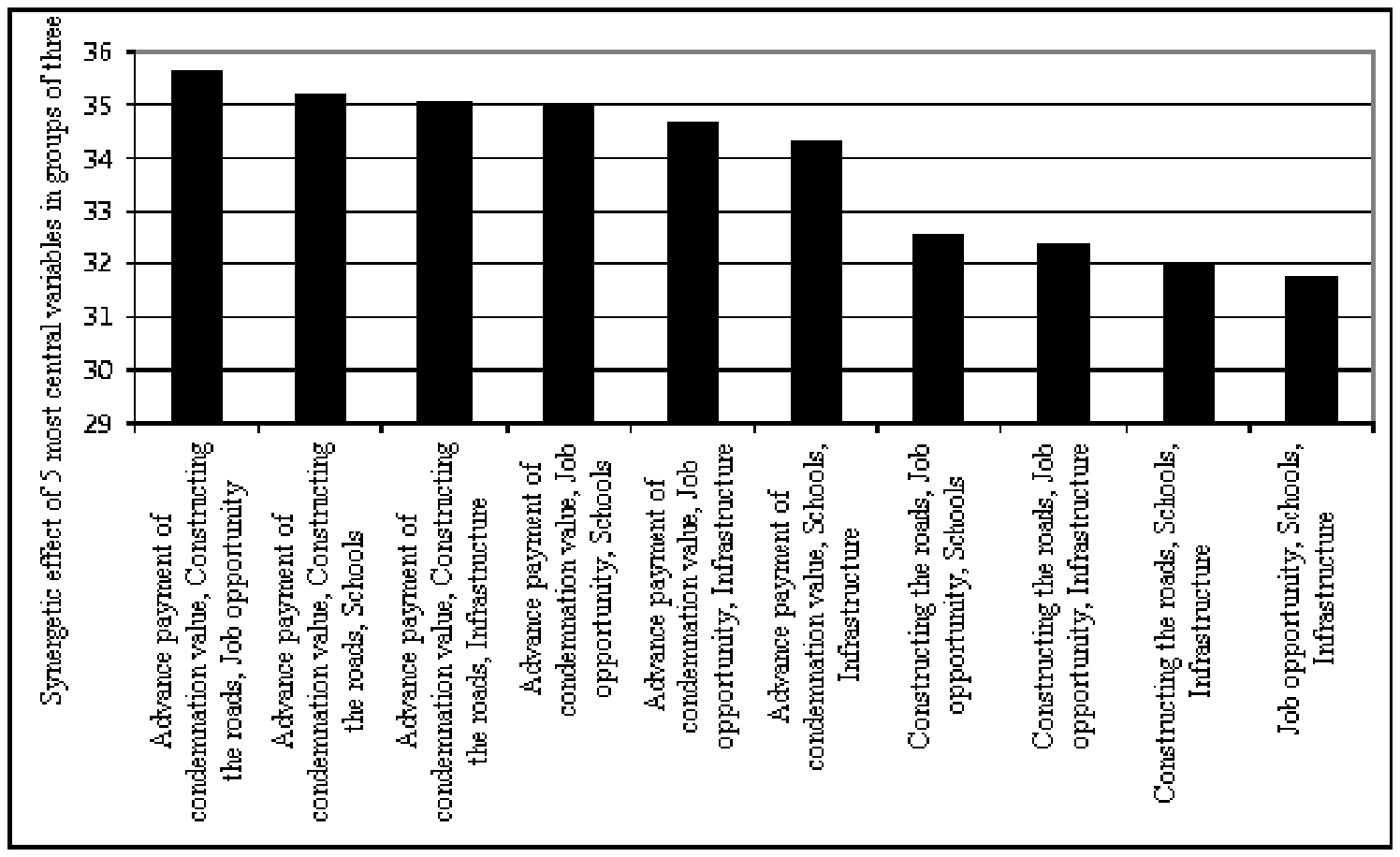}
\caption{The synergetic effect of the most central variables in triples causing a relative increase in the welfare of the people impacted by the dam construction.}
\end{center}
\end{figure}

\clearpage

\begin{figure}[ht]
\begin{center}
\includegraphics[width=\textwidth]{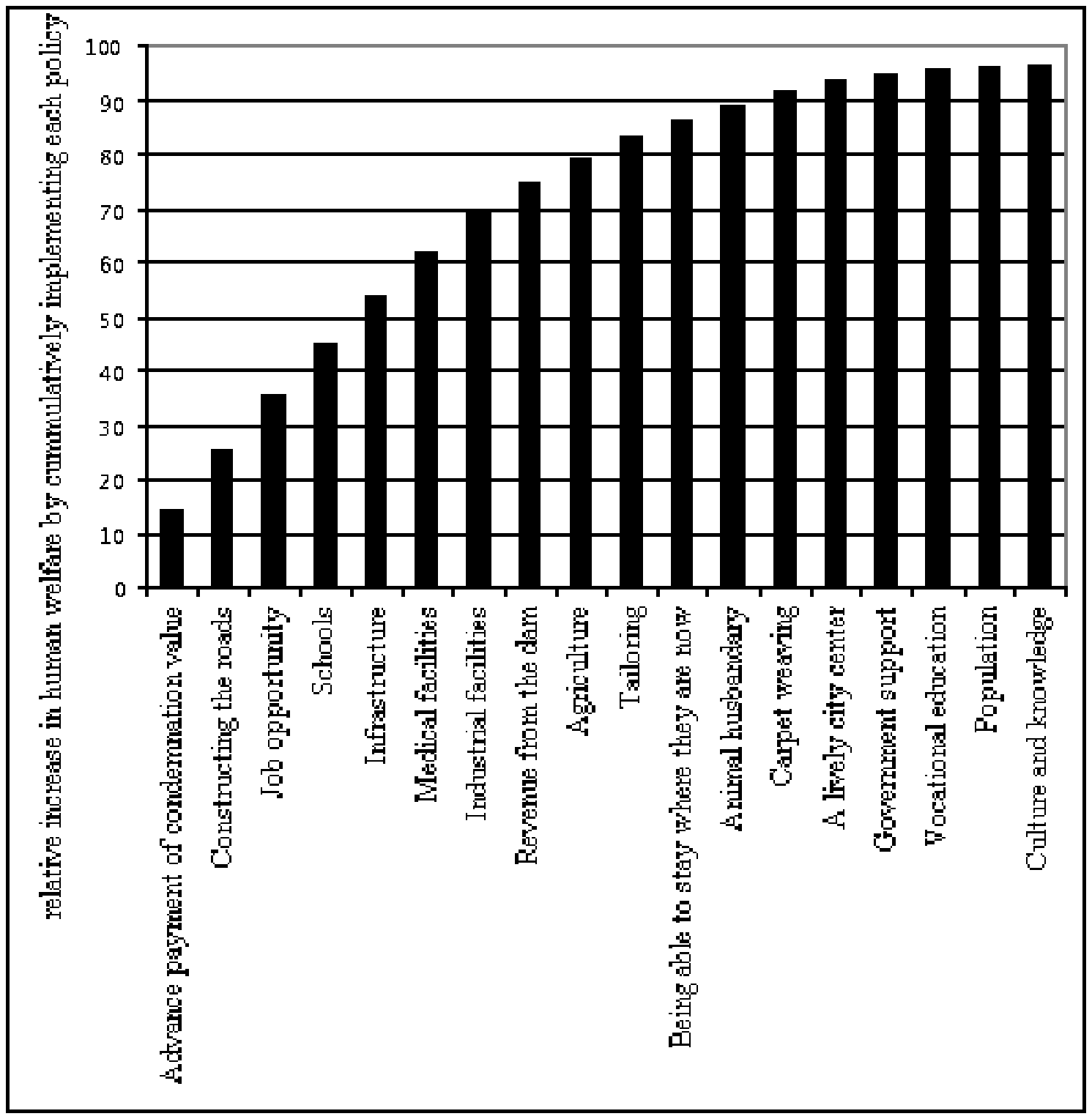}
\caption{The effect of the most central variables causing a relative increase in the income of the people impacted by the dam construction.}
\end{center}
\end{figure}

\end{document}